\documentclass{JINST}

\usepackage{color}
\makeatletter
\color{black}
\let\default@color\current@color
\makeatother

\def\pcomma{$^,$}
\def\itep{$^{1}$}
\def\pgwu{$^2$}
\def\pkent{$^{3}$}
\def\pglsg{$^4$}
\def\pmainz{$^5$}
\def\pbonn{$^6$}
\def\pjinr{$^7$}
\def\ppavia{$^8$}
\def\plpi{$^{9}$}
\def\psaint{$^{10}$}
\def\ppaviaUni{$^{11}$}
\def\pbasel{$^{12}$}
\def\pedinb{$^{13}$}
\def\pinr{$^{14}$}
\def\psackv{$^{15}$}
\def\pregina{$^{16}$}
\def\pzagreb{$^{17}$}

\def\puma{$^{18}$}
\def\pjerus{$^{19}$}

\title{A new measurement of the neutron detection efficiency for the NaI Crystal Ball detector
  }

\author{
 \vspace*{0.1in}
\begin{center}
(A2 Collaboration at MAMI)
\end{center}
\vspace*{0.1in}
M.~Martemianov\itep,
V.~Kulikov\itep\thanks{Corresponding author, E-mail: kulikov@itep.ru}~,
B.~T.~Demissie\pgwu,
Z.~Marinides\pgwu,
C.~S.~Akondi\pkent,
J.~R.~M.~Annand\pglsg,
H.~J.~Arends\pmainz,
R.~Beck\pbonn,
N.~Borisov\pjinr,
A.~Braghieri\ppavia,
W.~J.~Briscoe\pgwu,
S.~Cherepnya\plpi,
C.~Collicott\pmainz\pcomma\psaint,
S.~Costanza\ppavia\pcomma\ppaviaUni,
E.~J.~Downie\pgwu\pcomma\pmainz,
M.~Dieterle\pbasel,
M.~I.~Ferretti Bondy\pmainz,
L.~V.~Fil'kov\plpi,
S.~Garni\pbasel,
D.~I.~Glazier\pglsg,
D.~Glowa\pedinb,
W.~Gradl\pmainz,
G.~Gurevich\pinr,
D.~Hornidge\psackv,
G.~M.~Huber\pregina,
A.~Kaeser\pbasel,
V.~L.~Kashevarov\pmainz,
I.~Keshelashvili\pbasel,
R.~Kondratiev\pinr,
M.~Korolija\pzagreb,
B.~Krusche\pbasel,
A.~Lazarev\pjinr,
J.~M.~Linturi\pmainz,
V.~Lisin\pinr,
K.~Livingston\pglsg,
I.~J.~D.~MacGregor\pglsg,
D.~M.~Manley\pkent,
P.~P.~Martel\pmainz\pcomma\psackv,
D.~G.~Middleton\pmainz\pcomma\psackv,
R.~Miskimen\puma,
A.~Mushkarenkov\puma,
A.~Neganov\pjinr,
A.~Neiser\pmainz,
M.~Oberle\pbasel,
M.~Ostrick\pmainz,
P.~Ott\pmainz,
P.~B.~Otte\pmainz,
B.~Oussena\pgwu\pcomma\pmainz,
P.~Pedroni\ppavia,
A.~Polonski\pinr,
S.~Prakhov\pmainz,
G.~Ron\pjerus,
T.~Rostomyan\pbasel,
A.~Sarty\psaint,
D.~M.~Schott\pgwu,
S.~Schumann\pmainz,
V.~Sokhoyan\pgwu\pcomma\pmainz,
O.~Steffen\pmainz,
I.~I.~Strakovsky\pgwu,
Th.~Strub\pbasel,
I.~Supek\pzagreb,
M.~Thiel\pmainz,
A.~Thomas\pmainz,
M.~Unverzagt\pmainz,
Yu.~A.~Usov\pjinr,
S.~Wagner\pmainz,
D.~P.~Watts\pedinb,
J.~Wettig\pmainz,
D.~Werthm\"uller\pbasel,
L.~Witthauer\pbasel
 ~and M.~Wolfes\pmainz
\\
\llap{\itep} Institute for Theoretical and Experimental Physics SRC ``Kurchatov Institute", Moscow 117218, Russia\\
\llap{\pgwu} The George Washington University, Washington, DC 20052-0001, USA\\
\llap{\pglsg} SUPA School of Physics and Astronomy, University of  Glasgow, Glasgow G12 8QQ, United Kingdom\\
\llap{\pmainz} Institut f\"ur Kernphysik, University of Mainz,
    D-55099 Mainz, Germany\\
\llap{\pbonn} Helmholtz-Institut f\"ur Strahlen- und Kernphysik,     University of Bonn, D-53115 Bonn, Germany\\
\llap{\pjinr} Joint Institute for Nuclear Research, 141990 Dubna, Russia\\
\llap{\ppavia} INFN Sezione di Pavia, I-27100 Pavia, Italy\\
\llap{\ppaviaUni} Dipartimento di Fisica, Universit\`{a} di Pavia, I-27100 Pavia, Italy\\
\llap{\plpi} Lebedev Physical Institute, 119991 Moscow, Russia\\
\llap{\psaint} Department of Astronomy and Physics,  Saint Mary's University, Halifax, Nova Scotia B3H 3C3, Canada\\
\llap{\pbasel} Department f\"ur Physik, University of Basel, CH-4056 Basel, Switzerland\\
\llap{\pedinb} SUPA School of Physics, University of Edinburgh, Edinburgh  EH9 3JZ, United Kingdom\\
\llap{\pinr} Institute for Nuclear Research, 125047 Moscow, Russia\\
\llap{\psackv} Mount Allison University, Sackville, New Brunswick E4L 1E6, Canada\\
\llap{\pregina} University of Regina, Regina, Saskatchewan S4S 0A2,
Canada\\
\llap{\pzagreb} Rudjer Boskovic Institute, HR-10000 Zagreb, Croatia\\
\llap{\pkent} Kent State University, Kent, Ohio 44242-0001, USA\\
\llap{\puma} University of Massachusetts, Amherst, Massachusetts 01003, USA\\
\llap{\pjerus} Racah Institute of Physics, Hebrew University of
Jerusalem, Jerusalem 91904, Israel
}

\abstract{We report on a measurement of the neutron detection
efficiency in NaI crystals in the Crystal Ball detector obtained
from a study of single $\pi^0$  photoproduction on deuterium
using the tagged photon beam at the Mainz Microtron. The results
were obtained up to a neutron energy of 400~MeV. They are compared
to previous measurements made more than 15 years ago at the pion
beam at the BNL AGS. }


\begin{document}

\section{Introduction}

The A2 Collaboration at MAMI  is engaged in a
program~\cite{Krusche11} to study $\gamma$n-interactions for which
neutron detection is of crucial importance. The A2 experiment
consists of several detector systems described below. Two of them,
the Crystal Ball (CB) and TAPS, have a good neutron detection
efficiency. The NaI CB detector was built originally at SLAC in the
mid 1970s  for use  with colliding  e$^+$e$^-$ beams at
SPEAR~\cite{Chan78}. Later the CB was involved in experiments at
DORIS and Brookhaven National Laboratory before arriving at MAMI
where it demonstrates an excellent performance for photon and
charged particle detection. Neutron detection with this detector has
been studied less.  The CB neutron detection efficiency was measured
previously in a 1997-1998 run at BNL~\cite{BNL01} using the reaction
$\pi^-  p \rightarrow \pi^0 n$. There are very few references in the
literature to neutron interactions in NaI. A short summary of this
topic can be found in~\cite{Alyea97}. Previous measurements are not
directly applicable to the current status of the CB because of
possible aging effects, differences in thresholds and in methods of
analyzing the data. In this paper, we describe a new study of the CB
neutron detection efficiency based on measurements of single $\pi^0$
photoproduction on a liquid deuterium target using the tagged photon
beam at MAMI.
\section{Experimental setup}
\begin{figure}[tbp] 
\centering
\vspace{-0cm}
\includegraphics[width=.8\textwidth]{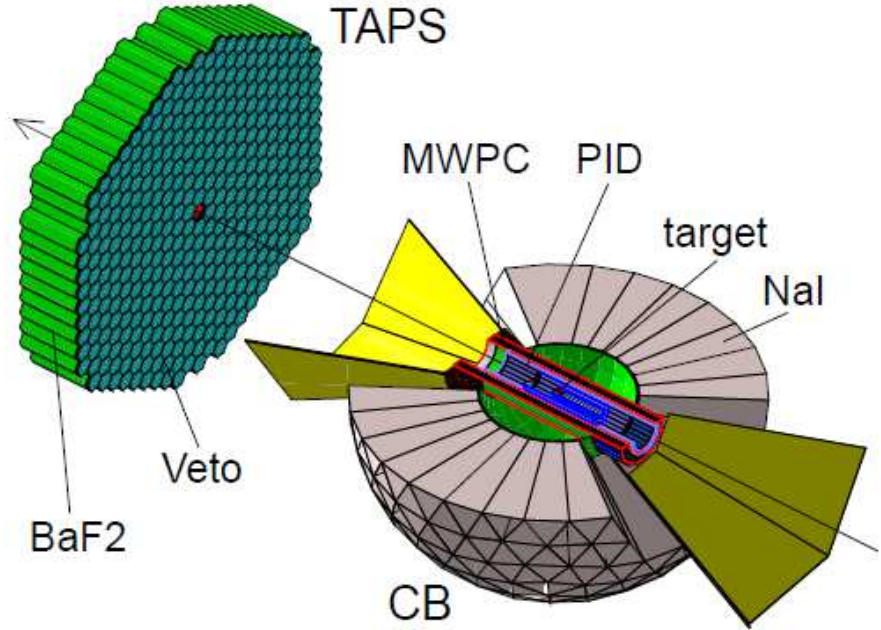}
\vspace{0cm} \caption{Layout of the A2 detector systems.}
\label{fig:Detector2}
\end{figure}

The measurements were performed at the tagged photon facility of the
Mainz Microtron accelerator (MAMI)~\cite{MAMI}. An electron beam was
used to produce bremsstrahlung photons, which were tagged with the
upgraded Glasgow magnetic spectrometer~\cite{Tagger}. The target was
a Kapton cylinder of 4~cm diameter and 10~cm length filled with
liquid deuterium. Photons, charged pions, and recoil nucleons,
produced in the target, were detected with an almost 4$\pi$
electromagnetic calorimeter schematically shown in
figure~\ref{fig:Detector2}. It combined the CB and the forward TAPS
detector \cite{TAPS}, although the latter was not used in the
current analysis.

The CB detector is a sphere consisting of 672 optically isolated
NaI(Tl) crystals, shaped as truncated triangular pyramids, which
point toward the center of the sphere. The crystals are arranged in
two hemispheres that cover 93\% of 4$\pi$ sr or polar angles of
20-160$^\circ$, sitting outside a central spherical cavity with a
radius of 25~cm. This cavity holds the target and inner
detectors. The NaI crystals have a length of 40.7~cm, which is equal
to 15.7~radiation lengths or $\sim$1 hadron interaction length. The
target was located in the center of the CB and was surrounded
by a Particle Identification Detector (PID)~\cite{PID} used to
distinguish between neutral and different charged  particles based
on $dE/dx$ measurements. It was made of 24 scintillator bars (50~cm
long, 4~mm thick) arranged as a horizontal cylinder with a radius of
6~cm. The PID was surrounded by two Cylindrical Multiwire
Proportional Chambers (MWPC). The inner (outer) chamber has a radius
of 7.4 (9.45)~cm and a length of 57~cm. Each chamber measured the
three-dimensional coordinates of a charged-particle track as a
result of a readout of three signals, one from a horizontal anode
sense wire and the other two from  spiral
strip cathodes~\cite{MWPC}.
\section{Overview of the method}

For the neutron detection efficiency measurement, we chose the
$\pi^0$
 photoproduction  on deuterium,
\begin{equation}
    \gamma + d \rightarrow \pi^0 + p + n,
\label{equ:1}
\end{equation}
where the $\pi^0$ decays to two photons. The reaction kinematics is
completely determined if the pion and proton  are detected and the
beam energy is known. The $\pi^0$ momentum is reconstructed using
the energies and directions of the decay photons. The proton
momentum is determined by its deposited energy in the CB and track
direction in the MWPC and the photon beam energy is given by the
tagging spectrometer. The reaction vertex was determined from the
intersection of the proton track with the photon beam axis
and allowed rejection of events from target walls.
 Then reaction (\ref{equ:1}) can be identified by selecting the neutron
by missing mass. This procedure rejects  events with more than one final-state pion and
provides a reconstruction of the neutron momentum vector. If neutron momentum
 points to the CB  (polar angles 30 - 150$^\circ$),
then a  neutral hit in the CB in the same direction is
searched for. Then the neutron detection efficiency is calculated as the ratio of
registered neutral hits to all neutrons.
 The indicated angular range is chosen to be smaller
 than the full acceptance of the CB to eliminate edge regions that
 inevitably have efficiency losses due to uncertainties in predicted and real hit positions.

Apart from being kinematically completely determined,  reaction
(\ref{equ:1}) has other advantages. It has a large cross section and
a broad energy and angle spectrum of neutrons. The energy spectrum
extends from nearly zero, when the pion is produced on a proton with
a spectator neutron, to the maximum energy from pion production on
the neutron. The latter was limited by the beam energy and the
available statistics. Reaction (\ref{equ:1}) was identified from
four clusters in the CB and one hit in the PID. Two clusters arise
from $\pi^0$ decay photons, the third one from  proton energy loss,
and the fourth from the neutron interaction with NaI. A cluster is
defined as a group of not more than 13 adjacent NaI crystals each
with deposited energy larger than 2~MeV centered around the crystal
with maximum energy deposition. The minimum cluster summed energy was
set to 15 MeV in the analysis.  In this measurement, we
selected events with three or four clusters in the CB and only one
hit in PID to reduce background from multiple pion production. In
the following sections, we describe all above mentioned steps in
more detail.

\section{Photon Tagger}

\begin{figure}[tbp] 
\centering
\begin{minipage}{0.49\linewidth}
\vspace{-.4cm}
\includegraphics[width=1.\textwidth]{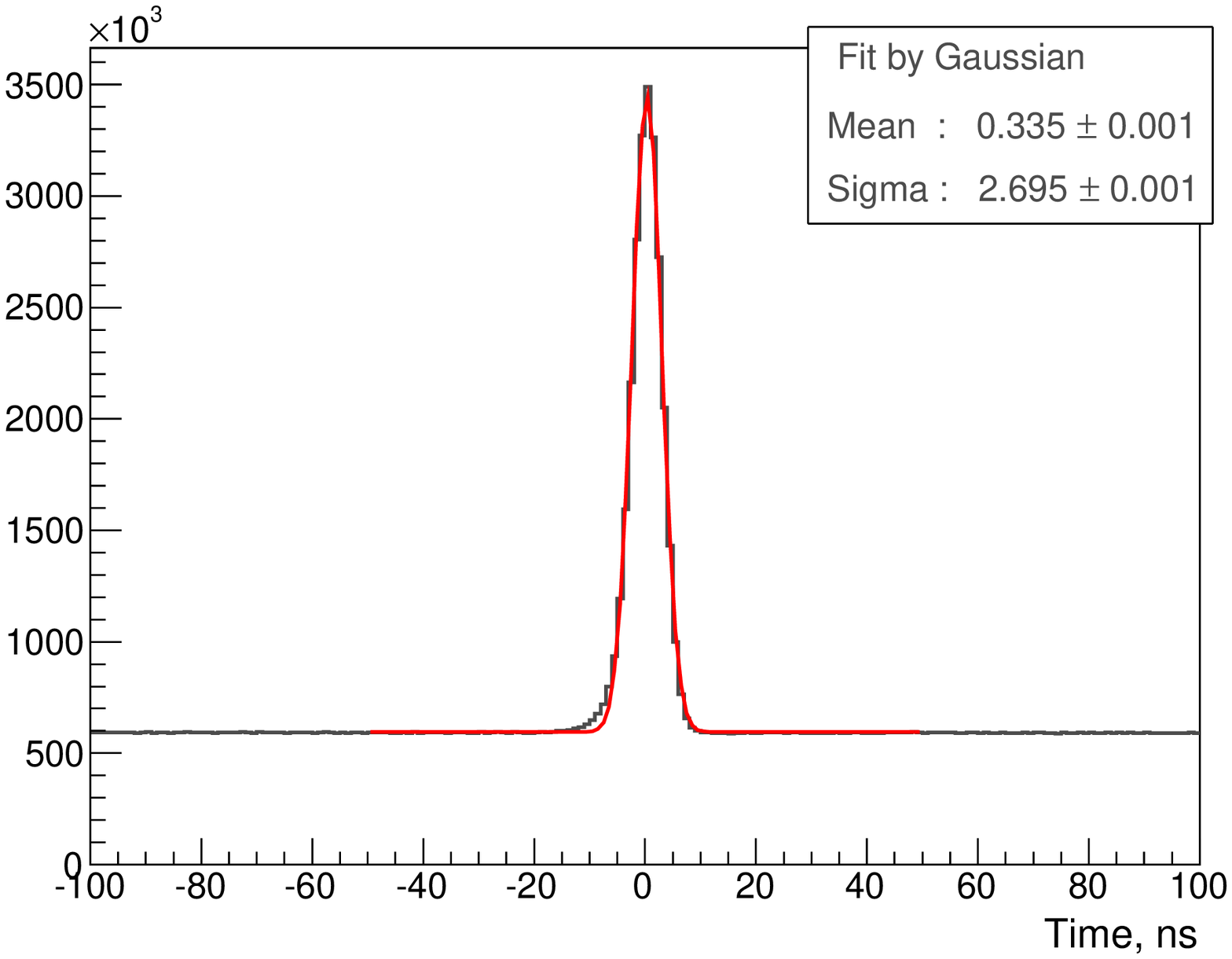}
\caption{Time distribution of tagged photons.}
\label{fig:TagTime}
 \end{minipage}
    \hfill
\begin{minipage}{0.49\linewidth}
\includegraphics[width=1.\textwidth]{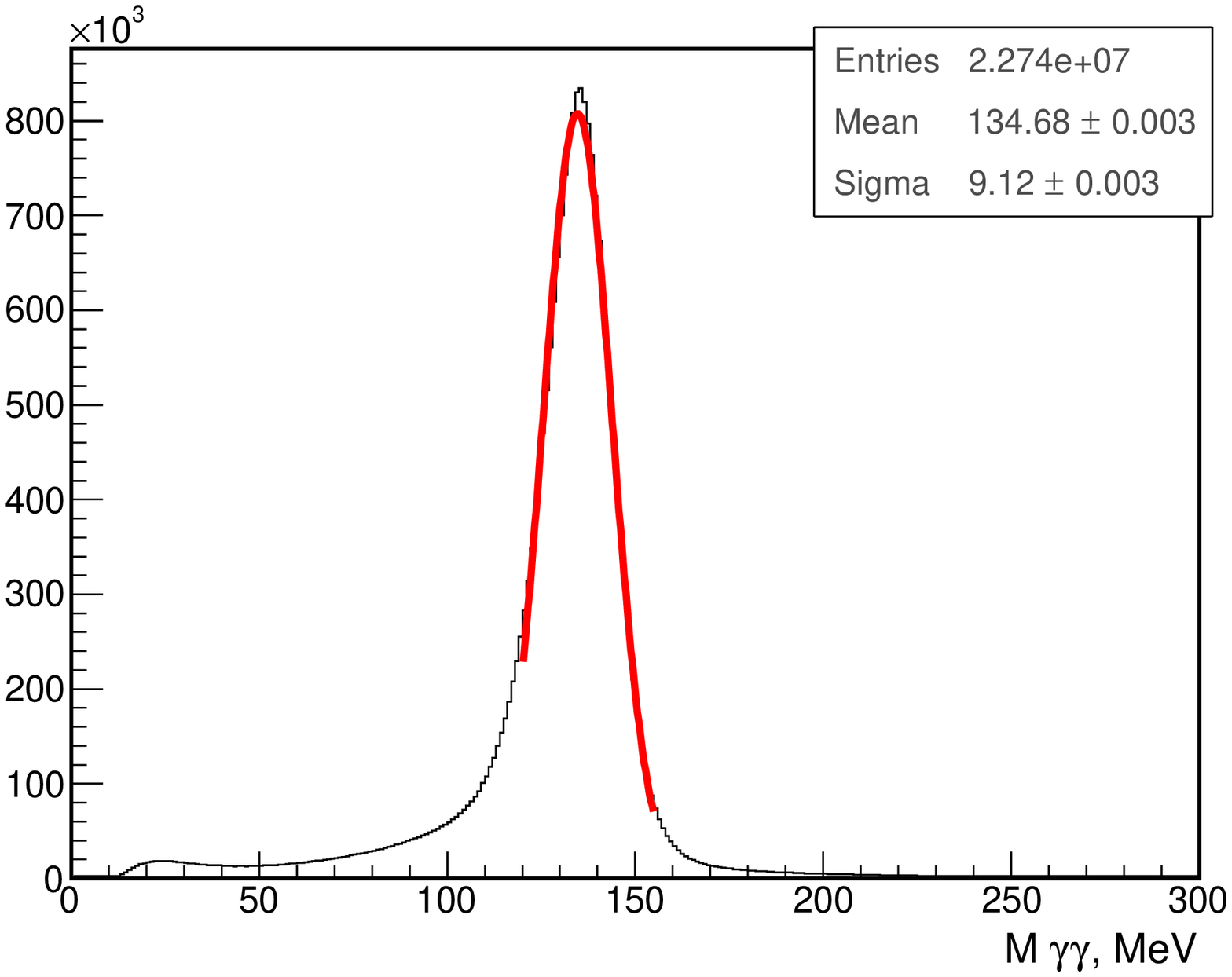}
\caption{Invariant $\gamma\gamma$-mass. A Gaussian fit to the peak of
the distribution is shown in red.} \label{fig:mpi0}
\end{minipage}
\end{figure}

For the neutron detection efficiency measurement, we used data from
the deuterium run taken at MAMI~\cite{MAMI} in March 2013.
Bremsstrahlung photons, produced by the 883~MeV electrons in a 10
$\mu$m Cu foil and collimated by a Pb collimator 3~mm in diameter,
were incident on a 10~cm long liquid deuterium target located at the
center of the CB. The beam   was ~1~cm in diameter at the target.
The energies of the incident photons were measured by detecting the
post-bremsstrahlung electrons in the 352 channel focal-plane
detector of the Glasgow-Mainz tagger~\cite{Tagger}. The energy
resolution of the tagged photons is determined mostly  by the width
of the tagger focal-plane detectors and by the electron-beam energy.
For the present beam energies, the typical width of a tagger channel
was about 2~MeV. The data were taken with a trigger that required
the total energy deposit in the CB to exceed 40~MeV.
Figure~\ref{fig:TagTime} shows the time distribution of tagged
photons relative to the trigger. The peak near 0~ns is due to the
prompt photons that produced the triggers. The entries  outside and
under this peak are due to random photons. The r.m.s. width of this
peak is equal to 2.7~ns and is determined mostly by the intrinsic
time resolution of the CB and time alignment of different channels.
More accurate adjustment of the channel time offsets can improve the
time resolution by a factor of two but it was not important for this
analysis. We selected the prompt photons in a time window  (-10~ns,
+10 ns) and subtracted background produced by random photons for all
distributions where the photon energy was used. The wide time range
accepted by the tagger made it possible to use wide time windows for
random photons (-230~ns to -30~ns and 30~ns to 230~ns) that resulted
in a negligible subtraction error.

\section{Selection of the single neutral pion}

\begin{figure}[tbp] 
\centering
\begin{minipage}{0.48\linewidth}
\vspace{-.4cm}

\hspace{-.5cm}
\includegraphics[width=1.1\textwidth]{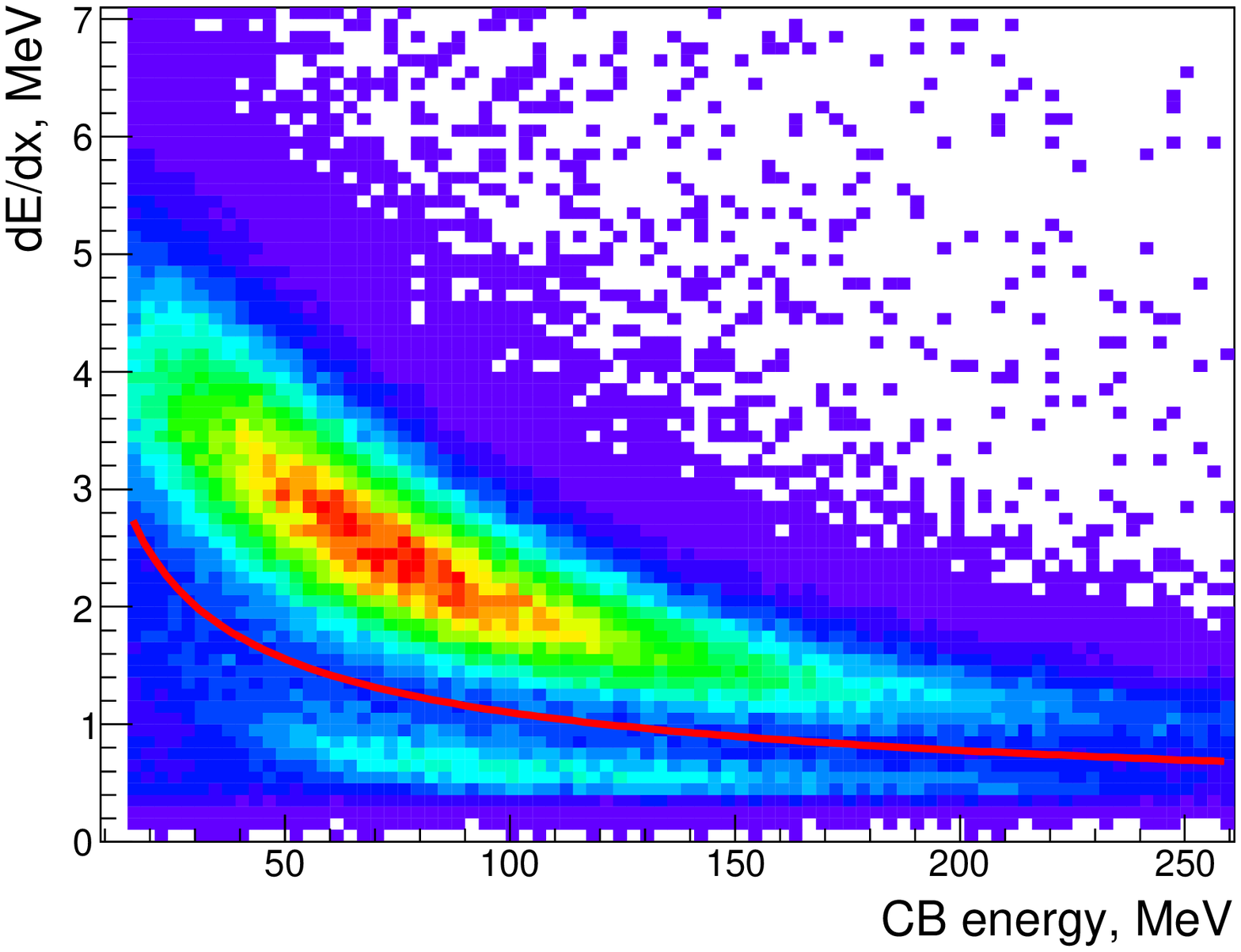}\flushleft
\flushright
\vspace{-.6cm}
\caption{ Pion/proton separation on the two dimensional plot $dE/dx$ as measured by PID vs.  CB total energy. Protons lie above the  empirically drawn line, pions - below.}
\label{fig:PIDvsCB}
\end{minipage}
   \hfill
\begin{minipage}{0.48\linewidth}
\flushleft
\hspace{-0.8cm}
\includegraphics[width=1.1\textwidth]{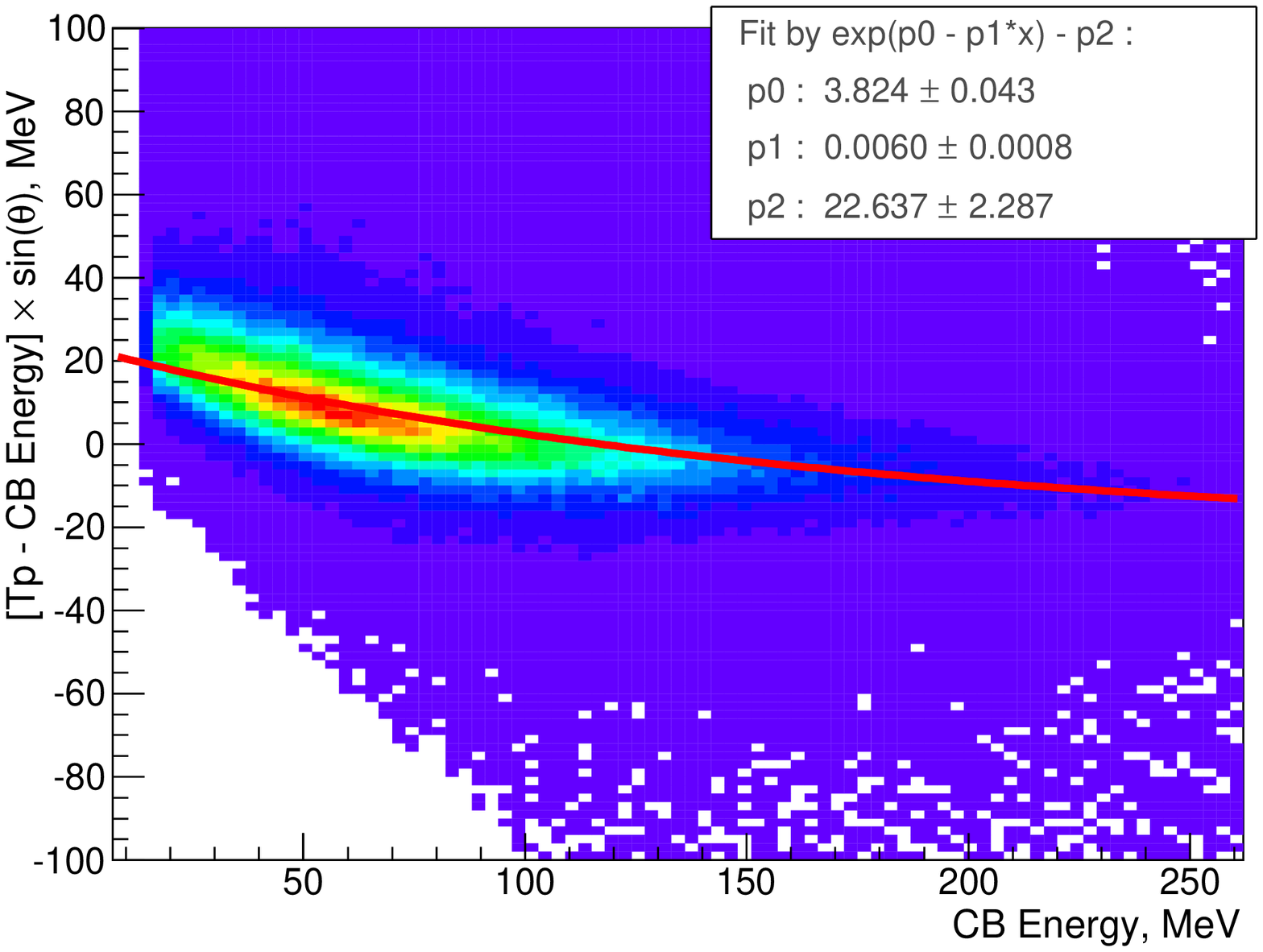}\flushleft
\vspace{-.6cm}
\caption{Proton energy correction for ionization losses and NaI
    light yield. $T_\textrm{p}$ - proton kinetic energy at interaction point,    $E$ - energy measured in the CB detector, $\theta$ - polar angle of    proton track. Line - fitted correction curve.}
    \label{fig:Proton Energy correction}
\end{minipage}
\end{figure}

The first step for reconstruction of  reaction (\ref{equ:1}) was
neutral pion selection. This was done by choosing the photon pair,
measured in the CB, with an invariant mass $m_{\gamma\gamma}$
closest to the $\pi^0$ mass $m_{\pi^0}$. An additional cut on the
sum of energies of the two photons $E_{\gamma\gamma} > m_{\pi^0}$
rejects low-energy photons. The $m_{\gamma\gamma}$ distribution
obtained with this cut is shown in figure~\ref{fig:mpi0}  with a
gaussian fit to the peak. The distribution demonstrates the excellent resolution
of the NaI CB detector in $m_{\pi^0}$-mass of ~9~MeV (r.m.s.). For
$\pi^0$ selection, we used a  120 < $m_{\gamma\gamma}$ < 150~MeV
cut. Identification of the proton and measurement of its energy were
done in a few steps. The proton track was required to have hits in
the PID and the CB and space points in  two MWPC
compatible with a straight line that has a point of closest approach
less than 10~mm from the beam axis within the volume of
the target. The separation of protons  from pions and
electrons/positrons is displayed in
figure~\ref{fig:PIDvsCB} where the CB cluster energy  is plotted along
the horizontal axis and $dE/dx$ is plotted along the vertical axis.
$dE/dx = E_\mathrm{\scriptscriptstyle{PID}}\sin\theta$, where
$E_\mathrm{\scriptscriptstyle{PID}}$ is the energy deposited in the
PID and $\theta$ is the track angle with respect to the beam axis.
The factor $\sin\theta$ corrects the deposited energy for track
length in the PID scintillator segments  which
are parallel to the beam axis. Ionization losses for protons depend on
their energies. In figure~\ref{fig:PIDvsCB}, protons lie above the
empirically drawn line while pions and electrons/positrons lie
below.

\begin{figure}[tbp] 
\centering \vspace{-0.5cm}
\includegraphics[width=.8\textwidth]{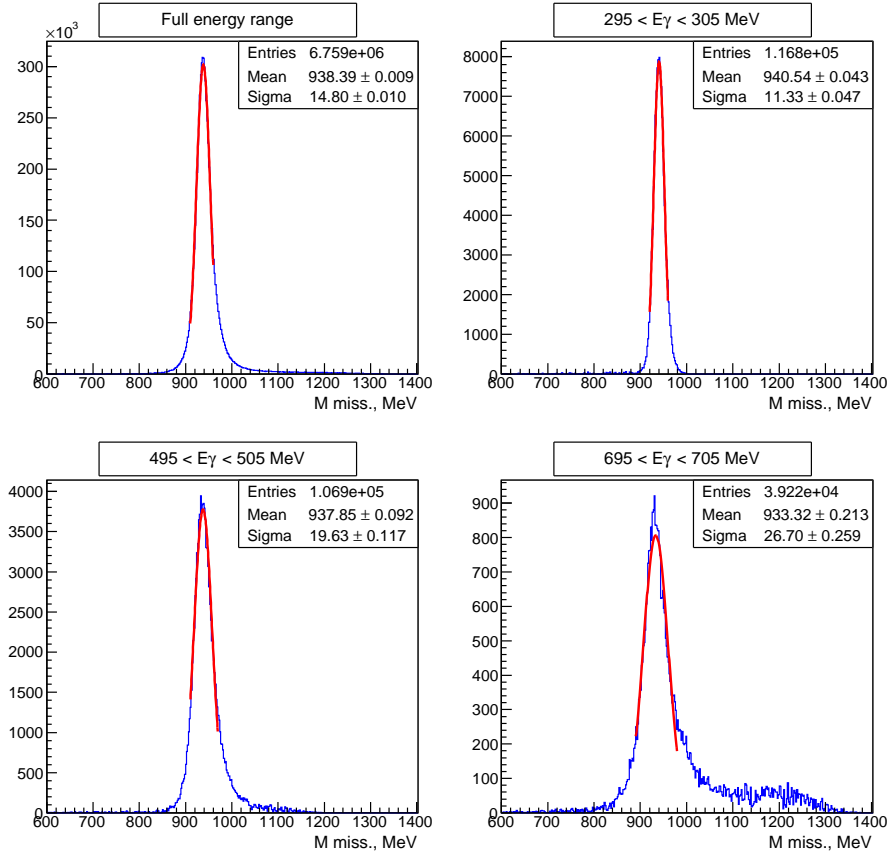}
\caption{ Missing mass distributions for the reaction $d(\gamma,\pi^0 p)X$ for different incident photon
    energies.  The peak position demonstrates good agreement with the neutron mass at all photon energies.} \label{fig:Correction quality}
\end{figure}

Two effects must  be taken into account to determine the proton
energy at the interaction point. First, due to  higher $dE/dx$ for
protons that are stopped in the NaI crystals of the CB detector, the
energy calibration for protons is different from that of minimum-ionizing
electromagnetic showers. Second, the measured proton energy
in the CB has to be corrected for ionization losses in  detector
materials between the interaction point and the CB crystals. These
materials are 2 cm of liquid deuterium, ~1~mm CH of the target
walls, 4~mm PID scintillator, and 1.5~mm stainless steel of the CB
spherical inner wall. The proton energy at the interaction point was
found by using the correction curve shown in figure~\ref{fig:Proton
Energy correction}, where $(T_\textrm{p} - E)\sin\theta$ is
plotted vs. the energy $(E)$ measured in the CB. Here $T_\textrm{p}$
is the proton kinetic energy at the interaction point calculated
from energy and momentum conservation for reaction (\ref{equ:1})
using photon and pion momentum vectors and the direction of the
proton track. At each bin of CB energy, $(T_\textrm{p} -
E)\sin\theta$ was fitted with a gaussian and the maximum was
found. Then these maxima at different energies were fitted with the
function $(\exp(c_{\scriptscriptstyle{1}} - c_{\scriptscriptstyle{2}}E) - c_{\scriptscriptstyle{3}})$ with
three free parameters $c_i, ~i=1-3$. The efficient performance of this
procedure is demonstrated in figure~\ref{fig:Correction quality},
where distributions of the events over missing mass
$M_\textrm{\small{miss}}$ in reaction (\ref{equ:1}) are given for a few
intervals of the beam energy. Missing mass is defined as
$M_\textrm{\small{miss}} = \sqrt{( P_{\gamma} + P_\textrm{d} - P_{\pi^0}
-P_\textrm{p})^2 }$, where $P_i ~(i=\gamma, d, \pi^0, p)$ are the
four-momentum vectors for the incident photon, target deuteron,
neutral pion, and proton, respectively. Good agreement between the
peak positions and the neutron mass is clearly seen in all four
photon energy ranges.

\section{CB neutron detection efficiency}

\begin{figure}[tbp] 
\centering
\begin{minipage}{0.42\linewidth}
\vspace{-0.4cm}
\hspace{-.4cm}
\includegraphics[width=1.1\textwidth]{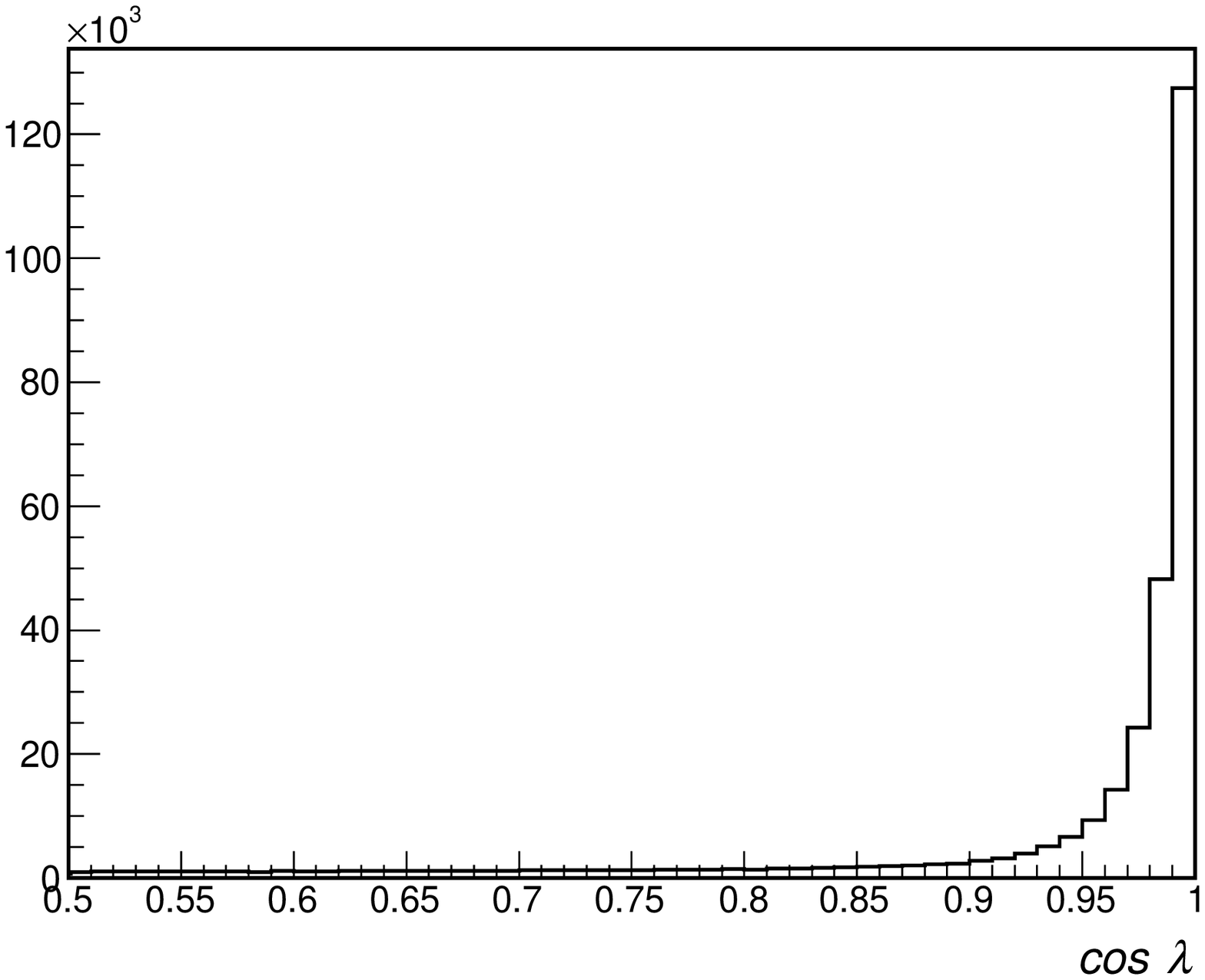}
\caption{Cosine of the angle between the predicted
    neutron momentum vector and the direction vector for third
neutral hit in CB.}
\label{fig:cosAngle}
\end{minipage}
\hspace{.2cm}
\begin{minipage}{0.42\linewidth}
\hspace{-.55cm}
\includegraphics[width=1.1\textwidth]{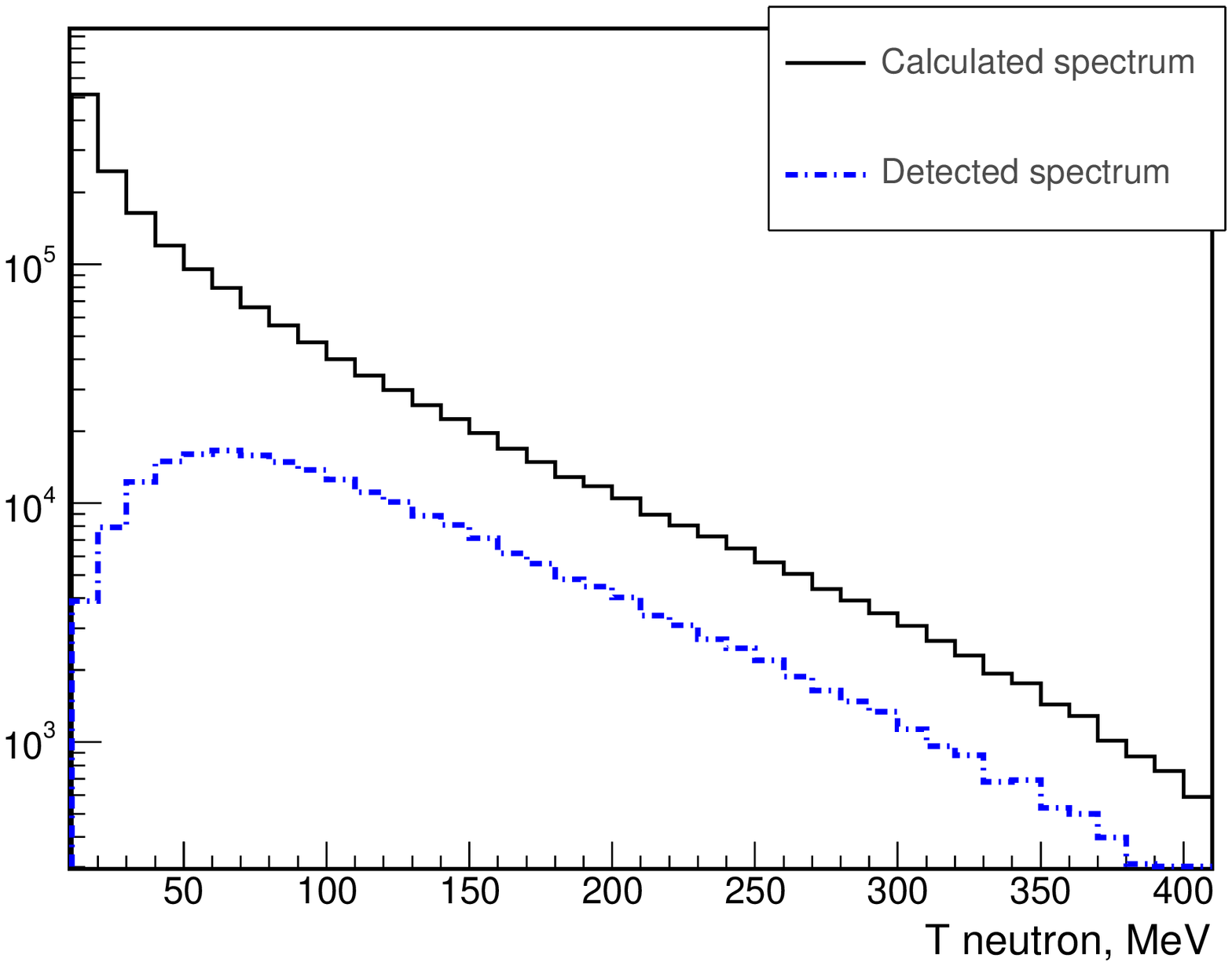}
\caption{Kinetic energy for neutrons identified by
    missing mass - solid histogram. When a neutron hit is registered the dashed histogram is obtained.} \label{fig:NEnergy}
\end{minipage}
\end{figure}

Neutrons were selected by a cut on the missing mass
$M_\mathrm{miss}$ centered at the mass of a neutron 890~<~
$M_\mathrm{miss}$~<~990~MeV. The momentum vector for a selected
neutron was calculated using energy and momentum conservation. This
vector must point  to a CB region slightly smaller than instrumented
with NaI crystals to eliminate edge effects; that is, the angle
$\theta$ of this vector with respect to the beam direction must be
$30^{\circ}<\theta<150^{\circ}$. The angles $\lambda$ between this
vector and the direction vectors for neutral hits in the CB were
determined. Hits from protons and photons from $\pi^0$ decay were
excluded from this procedure. The distribution over $\cos\lambda$ is
given in figure~\ref{fig:cosAngle}.
It has a large peak at $\lambda$ = 0 signifying the real neutron hit in the NaI crystals and a small  background of random hits. We chose $\cos\lambda$ = 0.85 as the boundary between the region of real neutron hits in the CB and background.
 The small uncertainty connected with this cut will be discussed later.
\begin{figure}[tp]  
\centering \vspace{-0.5cm}
\includegraphics[width=.8\textwidth]{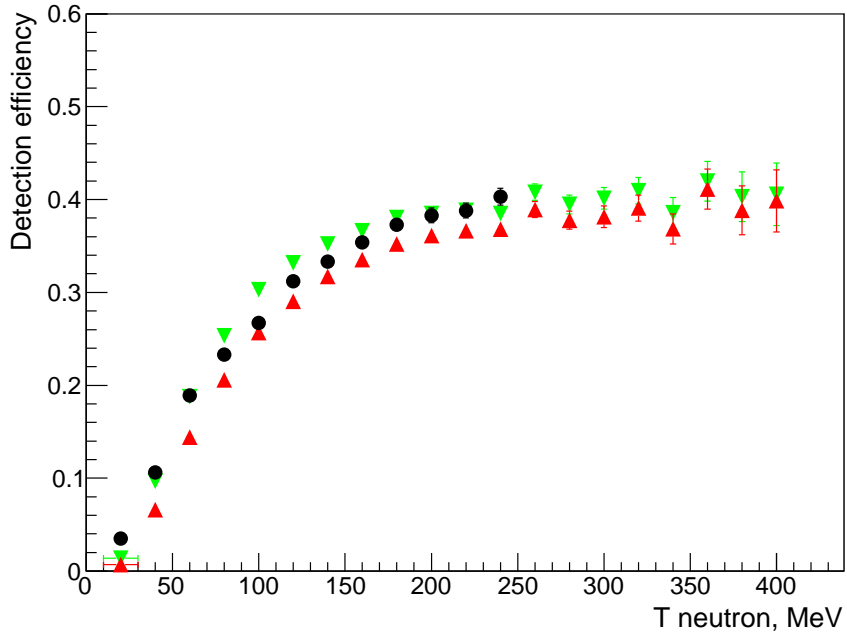}
\caption{CB detection efficiency  for neutrons as
a function of
    neutron kinetic energy. Downward (upward) triangles - present
    measurements for  a 15(20)~MeV cluster energy threshold;
    circles - previous measurement at BNL in 1997-98 for a
    20~MeV threshold \cite{BNL01}. } \label{fig:Efficiency}
\end{figure}

The energy distribution of neutrons from reaction (\ref{equ:1})
identified by missing mass  and incident on the CB
is shown by the solid line in figure~\ref{fig:NEnergy}.
Those neutrons that gave hits in the CB with $\cos\lambda$ < 0.85
have the energy distribution shown by the dashed line. The first
distribution is peaked at low energy, which demonstrates the large
contribution of spectator neutrons from pion photoproduction on the
proton. The second distribution goes to zero at low energies because
of the low detection efficiency for neutrons. The ratio of these two
distributions gives the CB detection efficiency for neutrons shown
in figure~\ref{fig:Efficiency} and listed in Table
\ref{tab:NEff2015} for 15 and 20~MeV cluster energy thresholds. The
detection efficiency is averaged over $\pm$ 10~MeV relative to the
listed energy values. The statistical and the two main systematic
uncertainties for 15~MeV threshold are also given. The statistical uncertainties
vary from less than 0.1\% at the lowest energy to 3\% at 400~MeV.
The increase results from the decrease in cross
section for reaction (\ref{equ:1}) as proton and neutron energy
increases. The systematic uncertainty I was estimated by changing the cut
on missing mass from (890 - 990) MeV to (890 - 1010)~MeV. The wider
window gives slightly lower detection efficiency due to a small
contamination from two-pion production. The smaller value of the
lower boundary of the cut has negligible effect on the neutron
detection efficiency. The systematic uncertainty II was estimated by
changing the cut on $\cos\lambda$ from the standard one
$\cos\lambda$ > 0.85 to $\cos\lambda$ > 0.65. Changing this cut
resulted in a higher detection efficiency as expected from the
larger acceptance for  neutron hits in the CB. Other cuts were also
tested: $\pi^0$ - mass, window for prompt photons, position of
interaction point in the liquid deuterium target, exclusion of dead
channels (1\% of all channels). Each of these cuts had an
order-of-magnitude smaller effect on neutron detection efficiency than I and
II, mentioned above. The total systematic uncertainty was obtained as the
quadrature sum of systematic uncertainties I and II and it is given in the
last column of Table \ref{tab:NEff2015}. For the most part, its
value is close to the statistical uncertainty. Table ~\ref{tab:NEff2015}
also demonstrates the sensitivity of neutron detection efficiency to
the cluster energy threshold.  An increase of the threshold from 15
to 20 MeV results in
 a relative decrease of neutron detection efficiency by 30\% at 40 MeV neutron energy
 and  by less than 5\% for  the  energies higher than 300 MeV.  In
figure~\ref{fig:Efficiency}, the CB neutron detection efficiency
measured more than 15 years ago ~\cite{BNL01} when the detector was
in operation at BNL is also shown. The shape of its energy
dependence for a 20~MeV threshold is in reasonable agreement with
the present measurement for a 15~MeV threshold. Absolute values of CB
neutron detection efficiency for the present measurement with a 20~MeV
threshold are 2-4\% smaller than measured at BNL.

\begin{table}[tbp]
\caption{\label{tab:NEff2015}  Neutron detection efficiency as a function of  neutron kinetic energy.
  Column 2 gives the efficiency
and statistical uncertainty for a 20 MeV threshold, column 3 gives the efficiency and statistical
uncertainty for~  a 15 MeV threshold, columns 4 and 5 give systematic uncertainties I and II, and
column 6 gives the total systematic uncertainty. See text for details.}
\begin{center}
\begin{tabular}{|c|c|c|c|c|c|c|c|c|c|}
\hline

Energy, & Efficiency, \%    & Efficiency, \%    & Sys.  & Sys.   &  Sys. \\
MeV   & (20 MeV threshold)  &  (15 MeV threshold)& I, \% & II, \%  & total, \%\\

 \hline
20  & 0.69$\pm$0.01  &  1.39$\pm$0.01  & 0.05  & 0.02  & 0.05 \\
40  &  6.6$\pm$0.1   &  9.7$\pm$0.1 & 0.1  & 0.1    & 0.2 \\
60  & 14.4$\pm$0.1   & 18.8$\pm$0.1 & 0.1  & 0.2   & 0.2 \\
80  & 20.6$\pm$0.2   & 25.3$\pm$0.2 & 0.1 & 0.3   & 0.3 \\
100 & 25.7$\pm$0.2   & 30.3$\pm$0.2 & 0.1  & 0.3  & 0.3 \\
120 & 29.0$\pm$0.3   & 33.1$\pm$0.3 & 0.2  & 0.3   & 0.3 \\
140 & 31.7$\pm$0.3   & 35.2$\pm$0.3 & 0.4 & 0.4  & 0.5 \\
160 & 33.5$\pm$0.4   & 36.6$\pm$0.4 & 0.4 & 0.4  & 0.6 \\
180 & 35.2$\pm$0.5   & 38.1$\pm$0.5 & 0.5  & 0.4  & 0.7 \\
200 & 36.1$\pm$0.5   & 38.5$\pm$0.5 & 0.5 & 0.5  & 0.7 \\
220 & 36.6$\pm$0.6   & 38.8$\pm$0.6 & 0.7 & 0.4  & 0.8 \\
240 & 36.8$\pm$0.7   & 38.5$\pm$0.7 & 0.6 & 0.4  & 0.7 \\
260 & 38.9$\pm$0.9   & 40.8$\pm$0.9 & 0.4 & 0.4  & 0.6 \\
280 & 37.8$\pm$1.0   & 39.5$\pm$1.0 & 0.7 & 0.7  & 1.0 \\
300 & 38.1$\pm$1.1   & 40.1$\pm$1.1 & 0.8 & 0.5  & 1.0 \\
320 & 39.1$\pm$1.4   & 41.0$\pm$1.2 & 0.6 & 0.7  & 1.0 \\
340 & 36.8$\pm$1.7   & 38.6$\pm$1.7 & 0.8 & 0.6  & 1.0 \\
360 & 41.1$\pm$2.1   & 42.0$\pm$2.1 & 0.5 & 0.4  & 0.6 \\
380 & 38.8$\pm$2.6   & 40.3$\pm$2.6 & 0.7 & 0.9  & 1.2 \\
400 & 39.8$\pm$3.4   & 40.6$\pm$3.4 & 0.4 & 0.3  & 0.7 \\

\hline
\end{tabular}
\end{center}
\end{table}
\section{Neutron interaction characteristics}

\begin{figure}[tbp] 
\centering

\begin{minipage}{0.48\linewidth}
\includegraphics[width=1.\textwidth]{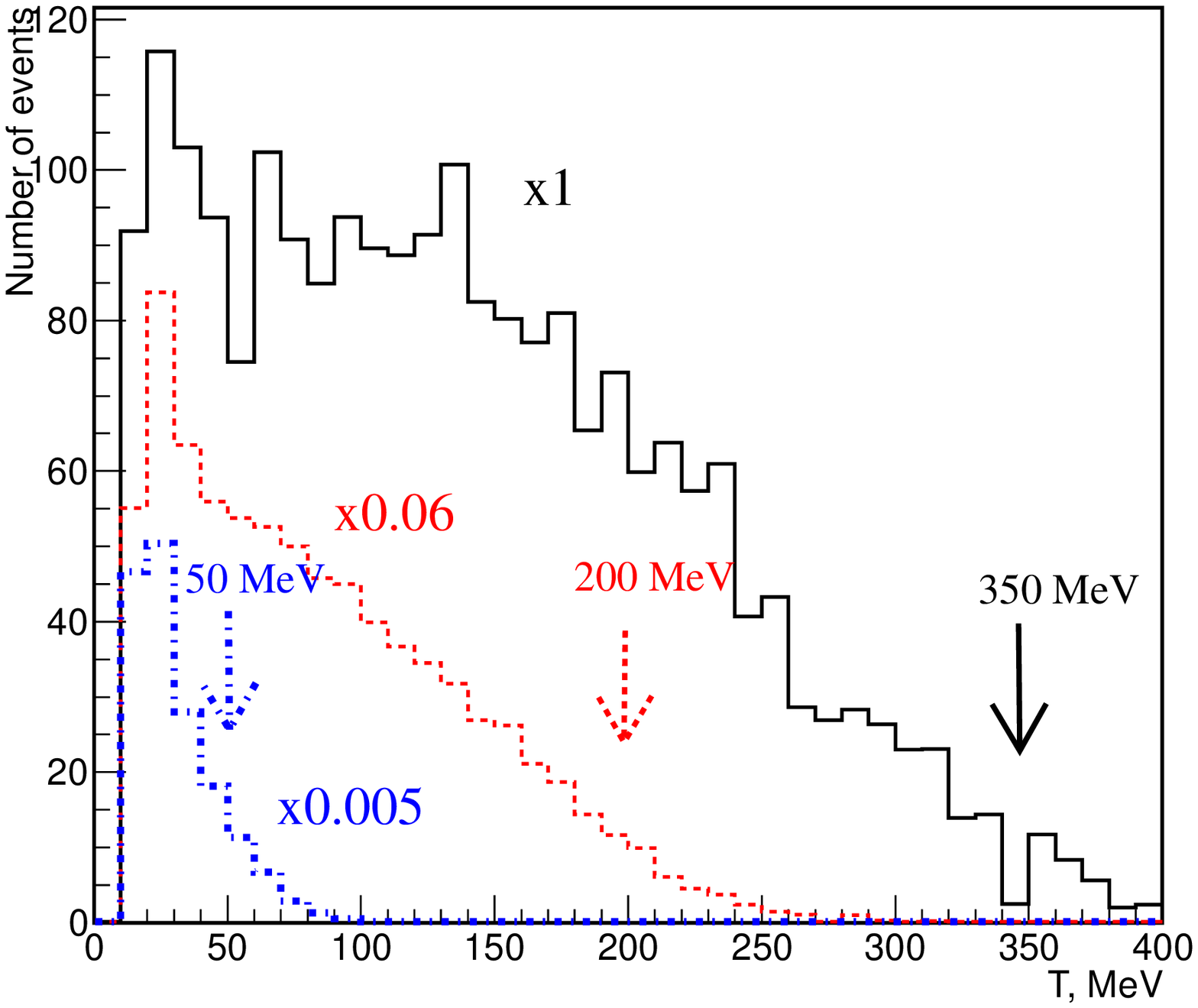}
\caption{Energy  deposited in the CB  for
    three incident neutron energies (marked by arrows): 350 $\pm$
    30~MeV - solid black line, 200 $\pm$ 20~MeV - dashed red line, 50    $\pm$ 10~MeV - dash-dot blue line. The histograms were multiplied   by the given factors to compensate for the strong energy dependence of the neutron flux.}
\label{fig:TnRegisted}
\end{minipage}
\hspace{.2cm}
 \hfill
\begin{minipage}{0.48\linewidth}
\vspace{-1.cm}
\includegraphics[width=1.\textwidth,height=.86\textwidth]{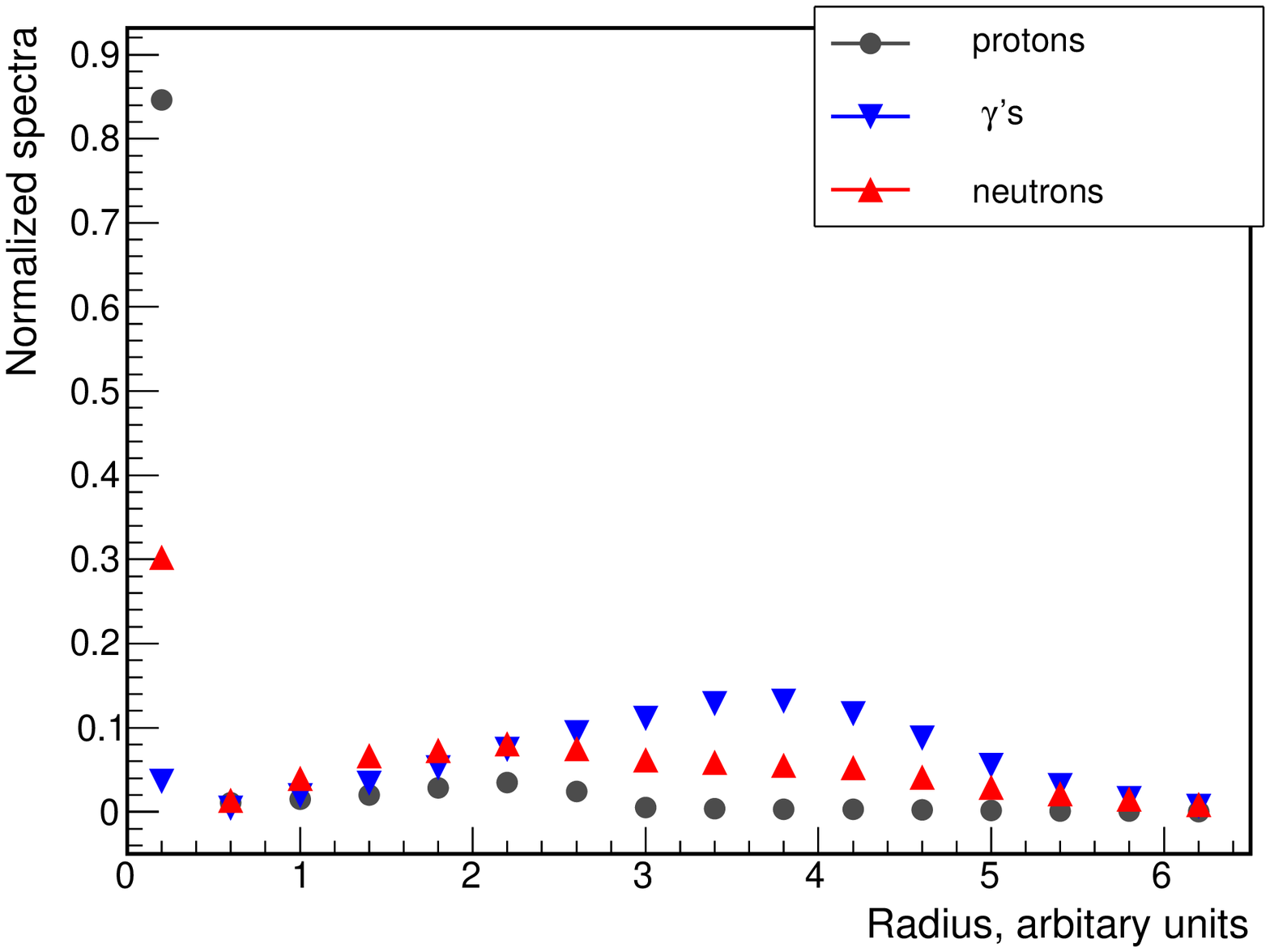}
\caption{Energy weighted ``radius" of the cluster for proton (black
    points), neutron (red upward triangles) and photon (blue downward triangles) induced   showers. The peak at zero is due to clusters that consist of only one crystal.} \label{fig:Radius}
\end{minipage}
\end{figure}

Neutrons interacting in the CB, which is not designed to be a
totally absorbing hadronic calorimeter, generally deposit only part
of their kinetic energy. This fact is illustrated by
figure~\ref{fig:TnRegisted} where distributions over cluster
energies recorded in the CB for hits identified as neutrons are
given for three selected intervals of  incident neutron energy: 50
$\pm$ 10 MeV, 200 $\pm$ 20~MeV, and 350 $\pm$ 30~MeV. For each
incident energy, the response in the CB is distributed from
threshold to a maximal energy that  slightly exceeds the incident
energy. This effect  can be attributed to energy resolution and to
some difference in calibration between hadron-induced and
photon-induced showers. The latter was used to determine the neutron
response. Figure~\ref{fig:TnRegisted} shows that the energy in a
cluster induced by a neutron  has limited utility for an estimation
of the neutron energy; however, the shape of a cluster can be more
informative. In figure~\ref{fig:Radius}, normalized distributions
over cluster ``radius" are given. The cluster ``radius" is an
energy-weighted sum of distances between the crystal with maximum
energy and each remaining crystal in a cluster. The cluster
``radius" is equal to zero for a cluster composed of only one
crystal. For a cluster induced by a neutron the probability to have
zero ``radius" is nearly 30\%. It is much higher than for a photon
cluster, for which the probability is only ~4\%. This difference can
be used to discriminate between neutron- and photon-induced showers.
Proton-induced showers have an even larger probability for zero
``radius" than neutron-induced showers. This information can also be
used to control a sample of protons.

\section{Discussion and conclusion}

We have measured the detection efficiency in the  NaI CB detector
for neutrons in the kinetic energy range  20 - 400~MeV using
 $\pi^0$ photoproduction on deuterium. A previous measurement
using the same detector was performed at BNL~\cite{BNL01} in
1997-1998 with the $\pi^-  p \rightarrow \pi^0 n$ reaction. The present
measurements agree closely with the previous measurements for the
shape of the energy dependence of neutron detection efficiency but
we find efficiencies that are 2-4\% smaller. This effect cannot be
unambiguously explained as a result of possible degradation of the
light yield of NaI crystals, which was part of the motivation for
this study. Many hardware and software changes have occurred between
these two measurements. Different energy calibration and clustering
algorithms can also result in a slightly different neutron
detection efficiency. The present measurement gives the up-to-date status of
neutron detection with the CB detector and can be used in various
physics analyses that involve neutrons and for improving the
description of neutron detection in Monte Carlo  simulations.

\section{Acknowledgements}

 The authors wish to acknowledge the outstanding support of the accelerator
group and operators of MAMI.
This work was supported by the Deutsche Forschungsgemeinschaft (SFB 443, SFB 1044), the INFN--Italy, the Schweizerischer Nationalfonds, the European Community-Research Infrastructure Activity under  FP7 programme (Hadron Physics2, grant agreement No. 227431), the UK Science and Technology Facilities Council (STFC 57071/1, 50727/1), the Natural Science and Engineering Research Council (NSERC) in Canada.
This material is based upon work
supported by the U.S. Department of Energy, Office of Science,
Office of Nuclear Physics, under Award Numbers DE-FG02-99-ER41110,
DE-FG02-88ER40415, DE-FG02-01-ER41194, and by the National Science
Foundation under Grant No. (PHY-1039130 \& IIA-1358175).
We thank the undergraduate students of Mount Allison University and The George Washington University for their assistance.
 M.M. and V.K. thank the Institut f\"ur Kernphysik at Mainz where part of this
work was performed for hospitality and support.



\begin{thebibliography}{9}

\bibitem{Krusche11} B.~Krusche, ~\emph{Photoproduction of mesons off
nuclei,} ~\emph{Eur.~Phys.~J.~ST} {\bf{198}} (2011) 199.


\bibitem{Chan78} Y.~Chan \textit{et al.}, ~\emph{Design and performance of a modularized NaI(Tl) detector (the Crystal Ball
prototype),} ~\emph{IEEE Trans.~Nucl.~Sci.} {\bf{25}} (1978) 333.

\bibitem{BNL01} T.~D.~S. Stanislaus \textit{et al.}, ~\emph{Measurement of neutron detection efficiemcies in NaI using Crystal Ball detector},
~\emph{Nucl.~Instrum.~Methods~A} {\bf{462}} (2001) 463.

\bibitem{Alyea97} J.~Alyea \textit{et al.}, ~\emph{Neutrons and the Crystal Ball Experiments}, ANL-HEP-TR-97-87.

\bibitem{MAMI} K.-H.~Kaiser \textit{et al.}, ~\emph{The 1.5 GeV harmonic double-sided microtron at Mainz University}, ~\emph{Nucl.~Instrum.~Methods~A} {\bf{593}} (2008) 159.

\bibitem{Tagger} J.~C.~McGeorge \textit{et al.}, ~\emph{Upgrade of
the Glasgow photon tagging spectrometer for Mainz MAMI-C},
~\emph{Eur.~Phys.~J.~A} {\bf{37}} (2008) 129.

\bibitem{TAPS} A.~R.~Gabler \textit{et al.}, ~\emph{Response of TAPS
to monochromatic photons with energies between 45-MeV and 790-MeV},
~\emph{Nucl.~Instrum.~Methods~A} {\bf{346}} (1994) 168.

\bibitem{PID} D.~Watts, ~\emph{The Crystal Ball and TAPS detectors at the MAMI electron beam facility}, ~\emph{Proceedings of the 11-th International Conference on Calorimetry in Particle Physics (Color2004), Perugia, Italy, 2004} ~(\emph{World Scientific, 2005}) 560.

\bibitem{MWPC} G.~Audit \textit{et al.},~\emph{DAPHNE: A Large acceptance tracking detector for the study of photoreactions at intermediate energies},~\emph{Nucl.~Instrum.~Methods~A} {\bf{301}} (1991) 473.

\end{thebibliography}
\end{document}